\documentclass[twocolumn,showpacs,eqsecnum,superscriptaddress,prb,citeautoscript,amsmath,amssymb,letterpaper]{revtex4}

\usepackage{graphicx}
\usepackage{multirow}
\usepackage{color}
\usepackage{bm}
\usepackage{times}
\usepackage{amsmath,bm,amsfonts}
\usepackage{dcolumn}
\usepackage{graphicx}
\usepackage{latexsym}
\usepackage{hhline}
\usepackage{hyperref}

\usepackage{braket}
\usepackage{bbold}

\newcommand{\ba}{\begin{eqnarray}}
\usepackage{manfnt}
\newcommand{\ea}{\end{eqnarray}}
\newcommand{\bd}{\begin{displaymath}}

\newcommand{\nn}{\nonumber \\}

\begin{document}
\title{Understanding resistance oscillation in CsV$_3$Sb$_5$ superconductor}
\author{Jung Hoon \surname{Han}}
\affiliation{Department of Physics, Sungkyunkwan University, Suwon 16419, South Korea}
\author{Patrick A. \surname{Lee}}
\affiliation{Department of Physics, Massachusetts Institute of Technology, Cambridge, Massachusetts 02139, USA}
\begin{abstract} A recent demonstration of the periodic oscillation of resistance in the thin film of CsV$_3$Sb$_5$ superconductor with a hole in the film suggests that charge-$4e$ and charge-$6e$ Cooper pairs may have condensed in this compound. While exciting, such interpretation calls for a precise determination of the effective area for the passage of Cooper pairs from one end of the lead to the other. Unlike the traditional Little-Parks effect where the rim around the hole is thin, the effective hole area  is not obviously defined for the ``thick-rim geometry'' adopted in Ref. \onlinecite{6e-wang}. Here, we note that the experiment was conducted in a regime where the superconctivity is strongly fluctuating, which motivates an analysis based on the spacetime formulation of the time-dependent Ginzburg-Landau theory. We argue that under appropriate conditions, the optimal semi-classical path is not the geometrically shortest one, but the one that moves along the edge of the hole and takes advantage of the reduced fluctuations at the boundary of the hole. The condition for the cross-over from the geometrically shortest path to the path that sticks to the wall is clarified. In such a scenario, the geometric area of the hole indeed emerges as the effective area for the flux, providing a  theoretical justification to the interpretation given in Ref. \onlinecite{6e-wang}. The conclusion of our analysis may have implication for similar experiments in other superconductors where the geometry of the device is not obviously that of the Little-Parks experiment employing the thin wall, but of the thick-rim type such as used in Ref. \onlinecite{6e-wang}. 
\end{abstract}
\date{\today}
\maketitle

\section{Introduction}
%\lee{Patrick, your writing will appear in blue color.} 
An explosion of activity has taken place around the discovery of vanadium-based kagome metals and the superconducting phase in one of the compounds, CsV$_3$Sb$_5$ (CVS)~\cite{wilson19,wilson20}. The multiple Fermi pocket structure in the normal state of CVS compound raises the possibility of pair density wave (PDW) - superconducting orders at a finite momentum ${\bf Q}$ vector, as recently detected experimentally~\cite{chen21,zqwang}. Furthermore, for symmetry reasons, PDW's at several ${\bf Q}$ vectors may appear simultaneously. As analylzed some time ago~\cite{agterberg08,berg09,radzihovsky09,agterberg11}, it is possible that above some temperature the PDW loses long range order, but a composite order $\Delta$ made up of the product of several primary PDW at different ${\bf Q}$ vectors according to 
\begin{align}
\Delta = \prod_{i=1}^N \Delta_{{\bf Q}_i } \label{eq:3Q}
\end{align}
may survive as the dominant order. Each ${\bf Q}_i$ here is the finite-momentum pairing field in the PDW phase and the order parameter $\Delta$ carries momentum  $\sum_i {\bf Q}_i$. The effective charge associated with $\Delta$ is $N(2e)$. In the case of kagome metal, the likely scenario is the condensation of three ($N=3$) ${\bf Q}$ vectors related by 120$^\circ$ forming a charge-$6e$ `Cooper molecule' with total momentum $\sum_i {\bf Q}_i = 0$. One telltale demonstration of such Cooper molecule phase will be the resistance oscillation near or slightly above the critical temperature in the spirit of the Little-Parks experiment with flux periods of $h/(2Ne)$~\cite{little-parks-62,little-parks-64}. A recent demonstration of resistance oscillation in the CVS superconductor with flux periods going from $h/2e$ to $h/4e$ to $h/6e$~\cite{6e-wang} upon increasing the temperature is evidence that this scenario may have already been realized. 

\begin{figure}[tb]
\includegraphics[width=0.4\textwidth]{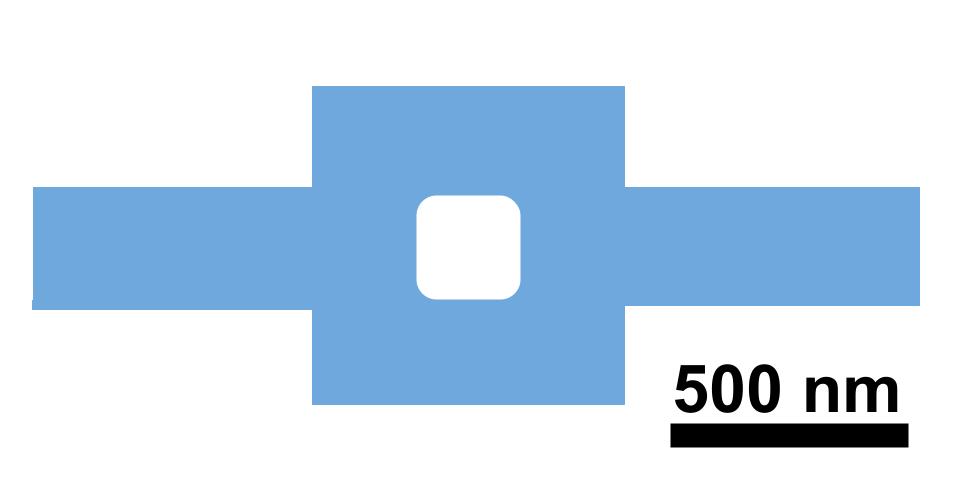}
\caption{Schematic geometry of the device used in the experiment of Ref. \onlinecite{6e-wang}. The white region is the hole area. Magnetic field is perpendicular to the plane.}   
\label{fig:0}
\end{figure}

In the original Little-Parks experiment, a thin cylinder with well-defined radius was used to measure the resistance oscillation near the critical temperature and the results were interpreted as periodic variations of the free energy as a function of flux quantum $h/2e$. The penetration depth $\lambda$ in the Little-Parks setup is considered much larger than the thickness of the cylinder because of vanishing superconducting order. We may call this the 'thin-rim geometry'. In Ref. \onlinecite{6e-wang} a hole was created at the center of the thin-film device similar to the Little-Parks, but the conducting region is not a thin cylinder of well-defined radius, as can be seen in the schematic reproduction of the experimental setup in Fig. \ref{fig:0}. This situation may be dubbed the 'thick-rim geometry'. 

Furthermore, the  resistance oscillations with flux periods of $h/4e$ and $h/6e$ were observed at temperatures well above the zero-resistance temperature $T_c$ where the magnetic field is expected to fully penetrate the device.  Since the Cooper pair experiences a varying range of enclosed flux as it goes from one lead to the other in the thick-rim geometry of such device, how does one decide on the effective area for the flux quantum? To understand the conundrum, we refer to Ref. \onlinecite{kogan04} where the free energy has been calculated on the annular geometry with  different inner ($a$) and outer ($b$) radii. In the absence of any vortex (which is what we assume as well), the free energy oscillation is governed by an effective area (adapted from Eq. (39) of Ref. \onlinecite{kogan04})
\begin{align}
    A_{\rm eff} = \frac{\pi (b^2 - a^2) }{2 \ln (b/a)} , 
\end{align}
reducing to the well-known result $\pi a^2$ only in the $b \rightarrow a$ limit. In general this effective area is somewhere between $\pi a^2$ and $\pi b^2$. For example, if $b=1.3a$ ($b=2a$) the effective area is predicted to be 1.32 (2.16) times the area of the hole. This is the area that is expected to determine the period of the original Little-Park experiment. Thus it is somewhat surprising that in the experiment of Ref. \onlinecite{6e-wang} the effective area was deduced to be close to the area of the hole, in all three temperature regimes where the flux quantizations are found to be in  multiples of $h/2e$, $h/4e$, or $h/6e$. In fact, the estimated area was slightly less than the geometric area of the hole they have in the device. It seems that the use of such area is not well justified {\it a priori} given the large width of the conducting region. Vortices will not play a significant role in the experiment either. The thickness ($t$) of the devices used in Ref. \onlinecite{6e-wang} was in the range of 10-20 nanometers~\cite{6e-wang} while the penetration depth is essentially infinite. This is the situation studied by Pearl~\cite{pearl,kogan01} where the screening of the external magnetic field takes place over the Pearl length $\Lambda= 2\lambda^2/t \gg \lambda$, larger than the device dimension in Ref. \onlinecite{6e-wang}. (The $T=0$ estimate of the penetration depth is already in excess of 300 nm~\cite{duan21}.) For the problem under consideration, then, it is safely assumed that there is little effect of vortex fluctuation and the magnetic field is nearly uniform. If so, the question of a well-defined effective area for the flux quantization looms large. 

An important hint comes from the observation that in  the experiment of Ref. \onlinecite{6e-wang}  the periodicity of the oscillations is best defined at the highest temperatures, where the resistivity is about half of the normal state and the $h/6e$ flux quantum is observed. This places us firmly in the fluctuating superconducting regime, i.e., the charge $6e$ order is far from achieving long range order. The system is best treated as a metal with some superconducting fluctuations. In contrast, at low temperatures where the theory of Ref. \onlinecite{kogan04} is expected to apply, only a few poorly formed oscillations has been reported, so that the period cannot be accurately determined. This motivates us to consider the fluctuating superconducting regime as opposed to the ordered regime considered in Ref. \onlinecite{kogan04}.

It is the purpose of this paper to focus on the fluctuation regime and address these questions in the framework of 
linearized time-dependent Ginzburg-Landau (TDGL) theory~\cite{tinkham}. We do not address the origin or the stability of $2Ne$-charged objects, but regard the effective charge $6e$ as an input in the TDGL. Instead, what we address is a problem of general relevance in all superconducting devices that are thin enough to be in the Pearl regime and devoid of screening, at the same time being thick-rimmed to defy a naive definition of the effective area as the geometric area of the hole.

\section{Formulation}
\label{sec:formulation} 
The linearized time-dependent Ginzburg-Landau (TDGL) equation near $T_c$ is~\cite{tinkham}
\begin{align}
    \left[ \overline{\alpha} - \frac{\hbar^2}{2m} 
    ({\bm \nabla} - i e^* {\bf A})^2 \right] \Psi = - \overline{\alpha} \tau \frac{\partial \Psi}{\partial t}. \label{eq:TDGL} 
\end{align}
The effective charge $e^* = 6e$ is assumed in the TDGL. The effective mass $m$ refers to that of charge-$6e$ composite. The symbol $m^*$ will be reserved for another definition of mass - see Eq. (\ref{eq:m-star}) - that will play a crucial role on the analysis. The gap parameter $\overline{\alpha}$ is proportional to $T-T_c$ and related to the coherence length $\xi$ by
\begin{align} \overline{\alpha} = \frac{\hbar^2}{2m \xi^2}  \end{align} 
while the relaxation time $\tau$ is given by~\cite{tinkham}
\begin{align} \tau = \frac{\hbar}{8 k_B (T- T_c )} . \end{align}
In this paper we shall take $\tau$ as a phenomenological constant. The coherence length of the bulk CVS superconductor is estimated to be 20-40 nm~\cite{wilson20,xhchen21a,xhchen21b}, comparable to the thickness of the device used in Ref. \onlinecite{6e-wang}. 

The TDGL equation of Eq. (\ref{eq:TDGL}) can be solved in general form by means of path integral technique. First one takes out $\overline{\alpha}$ through the substitution 
\begin{align} \Psi  = e^{-t/\tau} \psi, \end{align} 
as the new $\psi$ satisfies a different TDGL equation
\begin{align}
    \frac{\partial \psi}{\partial t} & = -\frac{\xi^2}{\tau} (-i {\bm \nabla} - e^* {\bf A})^2 \psi  
    \nn & = -\frac{\hbar}{2m^*} (-i {\bm \nabla} - e^* {\bf A})^2 \psi . \label{eq:TDGL-2} 
\end{align}
This resembles the Schr\"{o}dinger equation in imaginary time, and the effective mass 
\begin{align} m^* \equiv \frac{\hbar \tau}{2 \xi^2} , \label{eq:m-star} \end{align} 
differing from the mass of the charge-$6e$ Cooper pair $m$, shows up. As mentioned in the introduction, vortex effects will be considered weak and negligible. 

Equation (\ref{eq:TDGL-2}) is a diffusion equation with the vector potential ${\bf A} \neq 0$. When ${\bf A} =0$, 
\begin{align}
      \frac{\partial \psi}{\partial t} = \frac{\hbar}{2m^*} {\bm \nabla}^2 \psi \equiv D {\bm \nabla}^2 \psi  
\label{eq:diff-eq}   
\end{align}
is solved by the diffusion kernel (in two dimensions) 
\begin{align}
    \psi ({\bf r} t) & = \frac{1}{4\pi D t} \exp \left( -\frac{{\bf r}^2}{4Dt} \right) , \nn 
    \langle x^2 \rangle & = \langle y^2 \rangle = 2Dt = \frac{\hbar t}{m^*} . 
\end{align}

For ${\bf A} \neq 0$, the partial differential equation (\ref{eq:TDGL-2}) is solved in general as
\begin{align}
& \psi ({\bf r} t)  = \int U({\bf r} t; {\bf r}_0 t_0 ) \psi ({\bf r}_0 t_0 ) d {\bf r}_0  
\end{align}
where $t_0$ is some earlier time to $t$, ${\bf r}_0$ covers the area of the device (excluding the hole). The propagator $U$ is given in the path integral form
\begin{align}
& U({\bf r} t; {\bf r}_0 t_0 ) \nn 
& = \int [{\cal D} {\bf r} ] \exp \left[ i \frac{2\pi}{\Phi^*_0}  \int_{{\bf r}_0}^{\bf r} {\bf A} \cdot d {\bf r} - \frac{m^*}{2\hbar} \int_{t_0}^{t} \dot{\bf r}^2 dt  \right]  ,
\end{align}
where $\Phi^*_0 = h/e^*$ is the flux quantum pertinent to our problem. The sum $[{\cal D} {\bf r} ]$ spans all the spacetime paths that begin at $({\bf r}_0 t_0 )$ and end at $({\bf r} t)$. The Aharonov-Bohm (AB) phase factor appears despite the fact that we are solving the diffusion (imaginary-time Schr\"{o}dinger) equation. It is a common feature of geometric terms such as the AB phase to appear as a complex phase even in the imaginary-time evolution of the path integral. 

The product of two wave functions at unequal times and positions is $\Psi^* ({\bf r}_1 t_1 ) \Psi ({\bf r}_2 t_2 )$ where
\begin{align}
\Psi ({ \bf r }_2 t_2 ) & = e^{-(t_2 \!-\! t_0 ) /\tau} \int  U ({\bf r}_2 t_2 ; {\bf r}_0 t_0 )  \psi ({\bf r}_0 t_0 ) d {\bf r}_0  \nn 
\Psi^* ({ \bf r }_1 t_1 ) & = e^{-(t_1 \!-\!t_0 ) /\tau} \int  U^* ({\bf r}_1 t_1 ; {\bf r}'_0 t_0 ) \psi^* ({\bf r}'_0 t_0 ) d {\bf r}'_0 . 
\end{align}
Taking the initial time $t_0 = t_1$ in both path integrals without loss of generality gives
\begin{align}
& \Psi ({\bf r}_2 t_2 )  \Psi^* ({\bf r}_1 t_1 ) = e^{-(t_2 - t_1 )/\tau} \times  \nn 
&  ~~~~~ \int U ({\bf r}_2 t_2 ; {\bf r}_0 t_1 )  \psi ({\bf r}_0 t_1 ) \psi^* ({\bf r}_1 t_1 )  d{\bf r}_0  . \label{eq:product-of-psi-s} 
\end{align}

The non-local conductivity $\sigma ({\bf r}_1 , {\bf r}_2 )$ where ${\bf r}_1$ and ${\bf r}_2$ are the two leads coupled to the superconducting device is obtained from the current-current correlation function $\langle {\bf J} ({\bf r}_1 t_1 ) \cdot {\bf J} ({\bf r}_2 t_2 ) \rangle$~\cite{glazman}. In turn, the current-current correlation function follows from the two-particle density matrix
\begin{align}
& \Gamma ({\bf r}'_1 t_1 , {\bf r}'_2 t_2 ; {\bf r}_1 t_1 , {\bf r}_2 t_2 )  \nn 
& ~~ = \langle \Psi^* ({\bf r}'_1 t_1 ) \Psi ({\bf r}_1 t_1 ) \Psi^* ({\bf r}'_2 t_2 ) \Psi ({\bf r}_2 t_2 ) \rangle
\end{align}
by taking spatial derivatives
\begin{align}
& \langle {\bf J} ({\bf r}_1 t_1 ) \cdot {\bf J} ({\bf r}_2 t_2 ) \rangle  \sim  \nn 
& ~~ ({\bf D}_1 \! -\! {\bf D}^*_{1'} ) \cdot ({\bf D}_2 \!-\!  {\bf D}^*_{2'} )    \Gamma ({\bf r}'_1 t_1 , {\bf r}'_2 t_2 ; {\bf r}_1 t_1 , {\bf r}_2 t_2 )  |_{{\bf r} = {\bf r}'} .  \label{eq:J-J-corr} 
\end{align}
The $\langle \cdots \rangle$ implies thermal and other averages. The covariant derivatives are 
\begin{align}
    {\bf D}_{1,2} & = {\bm \nabla}_{1,2} - i e^* {\bf A} ({\bf r}_{1,2} )  \nn   
    {\bf D}^*_{1,2} & = {\bm \nabla}_{1,2} + i e^* {\bf A} ({\bf r}_{1,2} ) . 
\end{align}
We can express the two-point correlation function, using Eq. (\ref{eq:product-of-psi-s}), 
\begin{widetext} 
\begin{align}
%& \psi^* ({\bf r}'_1 t_1 ) \psi ({\bf r}_2 t_2 )  = e^{-(t_2 - t_1 )/\tau } \int U({\bf r}_2 t_2 ; {\bf r}_0 t_1 ) \psi ({\bf r}_0 t_1 ) \psi^* ({\bf r}'_1 t_1 ) d {\bf r}_0 , \nn 
%
%& \psi ({\bf r}_1 t_1 ) \psi^* ({\bf r}'_2 t_2 ) = e^{-(t_2 - t_1 )/\tau } \int U^* ({\bf r}'_2 t_2 ; {\bf r}'_0 t_1 ) \psi^* ({\bf r}'_0 t_1 ) \psi ({\bf r}_1 t_1 ) d {\bf r}'_0 ,  \nn 
%
& \Psi^* ({\bf r}'_1 t_1 ) \Psi ({\bf r}_1 t_1 ) \Psi^* ({\bf r}'_2 t_2 ) \Psi ({\bf r}_2 t_2 ) \nn %
& ~~~ = e^{-2(t_2 - t_1 )/\tau } \int U({\bf r}_2 t_2 ; {\bf r}_0 t_1 )   U^* ({\bf r}'_2 t_2 ; {\bf r}'_0 t_1 ) \psi ({\bf r}_0 t_1 ) \psi^* ({\bf r}'_1 t_1 ) \psi^* ({\bf r}'_0 t_1 ) \psi ({\bf r}_1 t_1 ) d {\bf r}_0 d {\bf r}'_0 . 
\end{align}
This is an exact representation of the two-point correlation function based on the path-integral solution of TDGL. Eventually, one takes ${\bf r}_{1,2} \rightarrow {\bf r}'_{1,2}$ and the integrals over ${\bf r}_0$ and ${\bf r}'_0$ pick up the largest contributions from ${\bf r}_0 \approx {\bf r}'_1$ and ${\bf r}'_0 \approx {\bf r}_1$. Under the approximation the two-particle correlation function simplifies to
\begin{align} 
\Gamma ({\bf r}'_1 t_1 , {\bf r}'_2 t_2 ; {\bf r}_1 t_1 , {\bf r}_2 t_2 ) & \rightarrow e^{-2(t_2 - t_1 )/\tau } \langle U({\bf r}_2 t_2 ; {\bf r}'_1 t_1 )   U^* ({\bf r}'_2 t_2 ; {\bf r}_1 t_1 ) 
| \psi ({\bf r}'_1 t_1 ) |^2  | \psi ({\bf r}_1 t_1 ) |^2  \rangle  \nn 
& \propto e^{-2(t_2 - t_1 )/\tau } \langle U({\bf r}_2 t_2 ; {\bf r}'_1 t_1 )   U^* ({\bf r}'_2 t_2 ; {\bf r}_1 t_1 ) \rangle . 
\end{align}
\end{widetext}
The first propagator $U({\bf r}_2 t_2 ; {\bf r}'_1 t_1 )$ is the sum over all paths from $( {\bf r}'_1 t_1)$ to $( {\bf r}_2 t_2 )$ while the second propagator $U^* ({\bf r}'_2 t_2 ; {\bf r}_1 t_1 )$ is (complex conjugate of) the sum over the paths from $( {\bf r}_1 t_1 )$ to $( {\bf r}'_2 t_2 )$. In the next section, we analyze the two-particle correlator in the semi-classical approximation with the geometry of Fig. \ref{fig:0}, similar to the one adopted in Ref. \onlinecite{6e-wang}, in mind. 

\section{Optimal path in the punctured geometry}
\begin{figure}[tb]
\includegraphics[width=0.4\textwidth]{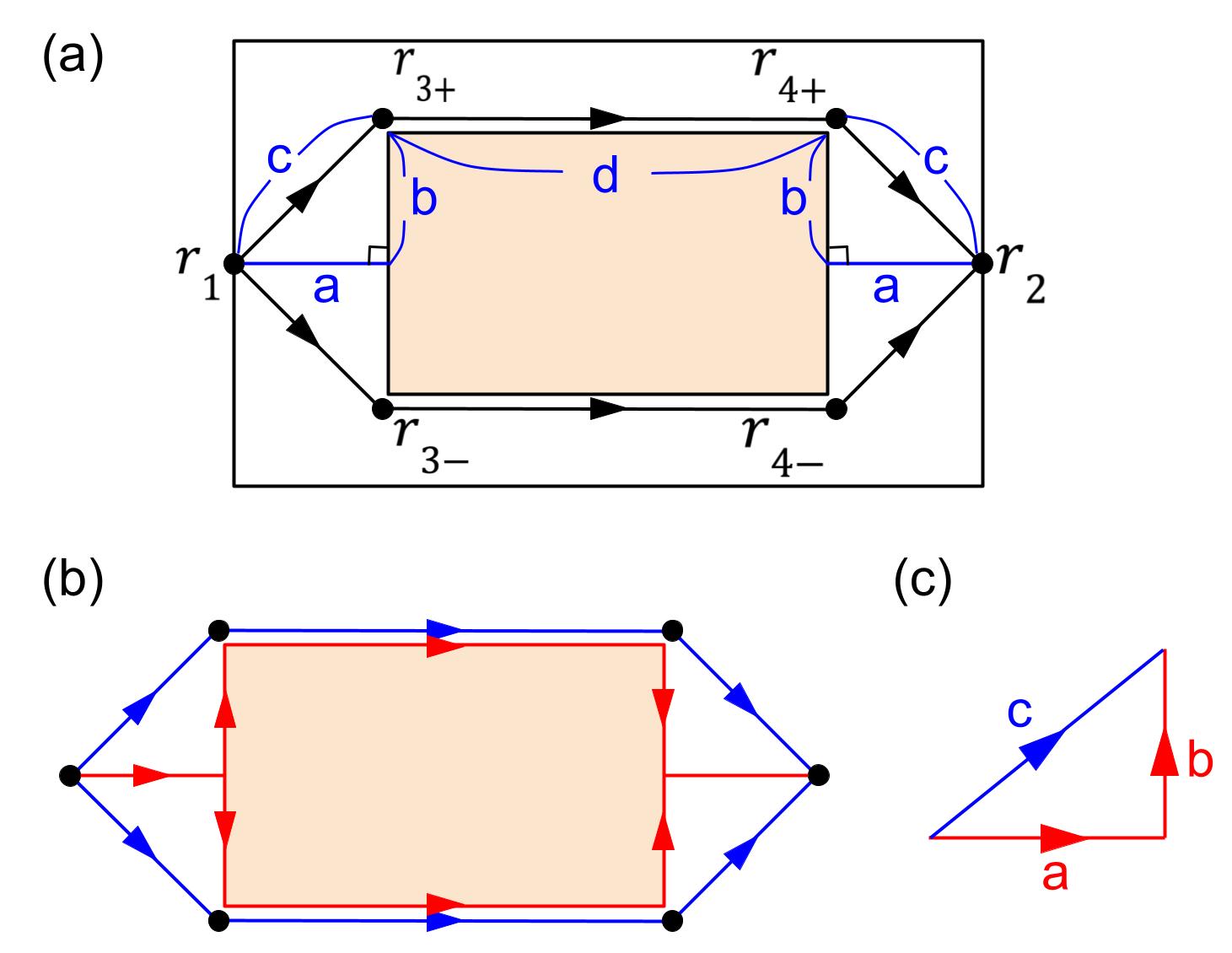}
\caption{(a) Schematic geometry of the device used in Ref. \onlinecite{6e-wang}. The leads shown in Fig. \ref{fig:0} are reduced to two point ${\bf r}_1$ and ${\bf r}_2$ now. The superconducting region is confined between the inner and the outer rectangles. The hole region is colored. Two geometrically shortest paths connecting the leads at ${\bf r}_1$ and ${\bf r}_2$ are shown as $({\bf r}_1 ; {\bf r}_{3\pm}; {\bf r}_{4\pm}; {\bf r}_2 )$. Lengths of various line segments are designated as $a, b, c, d$. (b) Geometrically shortest paths (in blue) and the alternative, edge Aharonov-Bohm path (in red). (c) The difference between blue and red paths is highlighted as the travel through the hypotenuse or the base + perpendicular sides of the right triangle.}  
\label{fig:paths}
\end{figure}

In the punctured geometry resembling the device of the experiment as shown in Fig. \ref{fig:paths}(a), paths can be divided as $+$ and $-$ depending on whether it passes over or under the inner square. There are four possibilities of paths for the product of propagators, $(++), (+-), (-+), (--)$. Keeping in mind that eventually we will take ${\bf r}'_{1,2} = {\bf r}_{1,2}$, the complex phase factors cancel out in the $(++)$ and $(--)$ cases but gives rise to the Aharonov-Bohm (AB) phase in the $(+-)$ and $(-+)$ cases. For the interpretation of the experiment in Ref. \onlinecite{6e-wang} it suffices to consider the two terms
\begin{align}
& \langle U({\bf r}_2 t_2 ; {\bf r}'_1 t_1 )   U^* ({\bf r}'_2 t_2 ; {\bf r}_1 t_1 ) \rangle \nn 
%& = \langle U_+ ({\bf r}_2 t_2 ; {\bf r}'_1 t_1 )   U^*_+  ({\bf r}'_2 t_2 ; {\bf r}_1 t_1 ) \rangle \nn 
%& + \langle U_- ({\bf r}_2 t_2 ; {\bf r}'_1 t_1 )   U^*_- ({\bf r}'_2 t_2 ; {\bf r}_1 t_1 ) \rangle \nn 
%
& \rightarrow \langle U_+ ({\bf r}_2 t_2 ; {\bf r}'_1 t_1 )   U^*_- ({\bf r}'_2 t_2 ; {\bf r}_1 t_1 ) \rangle \nn 
& ~~~ + \langle U_- ({\bf r}_2 t_2 ; {\bf r}'_1 t_1 )   U^*_+ ({\bf r}'_2 t_2 ; {\bf r}_1 t_1 ) \rangle . \label{eq:2p-correlator} 
\end{align} 

The propagators can be evaluated semi-classically. The exponential weight of the kinetic energy implies that shorter paths carry the larger weight. Taking ${\bf r}_1$ and ${\bf r}_2$ to be the opposite centers at the edges of the outer square, the shortest paths from ${\bf r}_1$ to ${\bf r}_2$ are the ones going just above and just below the inner square as shown in Fig. \ref{fig:paths}. The shortest paths are divided into three linear segments $({\bf r}_1 t_1 ; {\bf r}_{3\pm} t_3)$, $({\bf r}_{3\pm} t_3 ; {\bf r}_{4\pm} t_4)$, $({\bf r}_{4\pm} t_4 ; {\bf r}_2 t_2)$, and the corresponding classical action $S_{c} = (m^* /2) \int \dot{\bf r}^2 ~ dt$ evaluated for each:
\begin{align}
S_{13} = \frac{m^*}{2} \frac{c^2 }{t_3 \!- \!t_1} , S_{34} = \frac{m^* }{2} \frac{d^2}{t_4 \!-\! t_3} , S_{42} = \frac{m^* }{2} \frac{ c^2 }{t_2 \!-\! t_4 } .
\label{eq:bare-action} 
\end{align}
The total action $S = S_{13}+ S_{34} + S_{42}$ is minimized when we choose the time interval proportionately with distance, $t_3 - t_1 : t_4 - t_3 :  t_2 - t_4 = c : d : c$ and equals 
\begin{align} S ({\bf r}_1 , {\bf r}_2 ) = \frac{m^*}{2}\frac{(2c + d)^2}{t_2 - t_1} . \end{align} 

The semi-classical propagators are  
\begin{align}
& \langle U_\pm  ({\bf r}_2 t_2 ; {\bf r}'_1 t_1 )   U^*_\mp ({\bf r}'_2 t_2 ; {\bf r}_1 t_1 ) \rangle & \nn 
& \sim \langle \exp \left( i \frac{2\pi}{\Phi^*_0} \bigl[ \int_{{\bf r}'_1}^{{\bf r}_2} {\bf A} \cdot d{\bf r}_\pm - \int_{{\bf r}_1}^{{\bf r}'_2} {\bf A}\cdot d{\bf r}_\mp \bigr] \right) \nn 
& \times \exp \left( - S ( {\bf r}'_1 , {\bf r}_2 ) /\hbar - S ( {\bf r}_1 , {\bf r}'_2 )  /\hbar \right) \rangle . \label{eq:2-ptl-propagator}
\end{align}
Calculating the current-current correlation function using Eq. (\ref{eq:J-J-corr}) is straightforward and yields the oscillatory factor $\cos (2\pi \Phi_g /\Phi^*_0 )$ where $\Phi_g$ is the flux enclosed inside the geometric area spanned by the shortest path. This area is obviously larger than that of the inner square and therefore disagrees with an interpretation that associates the AB phase with the flux through the hole. Worse yet, the flux $\Phi_g$ depends explicitly on the geometric dimension of $a$ so that in a different device with the larger width $a$, a totally different estimate of $\Phi_g$ is inevitable. If such were the case, an interpretation of the resistance oscillation as due to flux periodicity is highly unreliable. 

The semi-classical path consistent with the conventional AB interpretation would be the one where both ${\bf r}_{3\pm}$ and ${\bf r}_{4\pm}$ hit the center of the inner square, and the remainder of the path goes around its perimeter as shown by red in Fig. \ref{fig:paths}(b). We refer to such alternative path as the edge Aharonov-Bohm (edge-AB) path since it goes around the edges of the hole and picks out its area. %In the remainder of the paper, the flux through the edge-AB path will be denoted as $\Phi_{AB}=BA$ where $A$ is the area of the hole . 
The difference between the edge-AB path and the geometrically preferred path is whether one goes from ${\bf r}_1$ to ${\bf r}_{3\pm}$ through the hypotenuse of the right triangle or through the base + perpendicular edges of it, as shown in Fig. \ref{fig:paths}(c). Designating the lengths of three edges as $a, b, c$, a simple back-of-the-envelop calculation shows that the action for the two paths (called $c$-path and $ab$-path) are respectively 
\begin{align}
    S_c & = \frac{m^*}{2}\frac{c^2}{t} \nn  
    S_a+ S_b & = \frac{m^*}{2}\frac{a^2 }{t'} + \frac{m^*}{2}\frac{b^2 }{t-t'} \nn 
    & \rightarrow \frac{m^*}{2} \frac{(a + b)^2 }{t} , \label{eq:Sabc}
\end{align}
the last line following upon minimizing $S_a + S_b$ with respect to the intermediate time $t'$. Since $(a+b)^2 > c^2$, we obtain $S_a + S_b > S_c$ as expected. This simple result is as expected, but poses a challenge in interpreting the experiment~\cite{6e-wang}. To get a heightened sense of the paradox, imagine the limit $a\rightarrow \infty$ such that the area enclosed by the $c$-path becomes very large. The flux enclosed inside the geometric area will vary rapidly with the magnetic field, and the oscillation will become unobservable. A more attractive scenario is that the particle somehow chooses the edge-AB path and maintain the same area for enclosing the flux quantum regardless of the size of $a$. But how? 

Below we propose a quick fix to the conundrum and discuss its physical origin in the following section. Instead of the bare action used in Eq. (\ref{eq:Sabc}), adopt the following effective action
\begin{align}
    S_a & = \frac{m^*}{2}\frac{a^2}{t'} + \alpha a^2 t' , \nn 
    S_b & = \frac{m^*}{2}\frac{b^2}{t-t'} + \beta b^2 (t-t') , \nn 
    S_c & = \frac{m^*}{2}\frac{c^2}{t} + \alpha c^2 t . \label{eq:Sabc0} 
\end{align}
The extra $\propto t$ term in the effective action can be derived in a standard manner as a result of fluctuation around the classical path (see following section). The coefficients $\alpha$ and $\beta$ are in principle different because in the case of $b$-path, the presence of the edge prohibits the fluctuation to roughly half the amount compared to $a$- or $c$-path where the fluctuation is uninhibited. For now we treat them as  parameters $\alpha, \beta > 0$ and minimize the new action $S_a + S_b$ with respect to $t'$, yielding  
\begin{align}
\frac{m^*}{2} \frac{a^2}{t'^2} - \frac{m^*}{2} \frac{b^2}{(t-t')^2}  =  \alpha a^2 - \beta b^2 . 
\end{align}
An approximate solution is found by parameterizing $t' = t [ a/(a+b) - \epsilon]$ and $t-t' = t[ b/(a+b) + \epsilon]$ and finding, to first order in $\epsilon$:

\begin{align}
    \epsilon \approx \frac{( \alpha a^2 - \beta b^2 ) a b}{ (a+b)^4} \frac{t^2}{m^*} . 
\end{align}
We get $\epsilon >0$ when $\alpha a^2 > \beta b^2$. This makes sense, as the particle wants to spend more time ($\epsilon >0$) in the $b$-region where the linear-in-$t$ part of the action is less than that in the $a$-region. To the same order in $\epsilon$ we find
\begin{align}
    S_a + S_b - S_c & \approx \frac{m^* ab}{t} - \left( \alpha ab + (\alpha -\beta) \frac{b^3}{a+b}\right) t   \nn 
    & - \frac{(\alpha a^2  - \beta b^2 )^2 ab}{(a+b)^4} \frac{t^3}{m^*} .
\label{eq:Sabc}
\end{align}

For $\alpha > \beta$ the second term which is linear in $t$ is negative and will dominate the first term for large enough $t$, making the $ab$-path more favorable than the $c$-path despite its longer trajectory. Similarly the third term is negative if $\beta$ is small enough compared with $\alpha$, only speeding up the crossover from $c$- to $ab$-path. We shall later show that in fact $\beta$ appraches zero. This will be our route to stabilizing the long path that hugs the edge. The expression derived in Eq. (\ref{eq:Sabc}) invites the definition of a time scale $t_c$:
\begin{align}
    t_c = \sqrt{m^*/ \alpha} . 
\end{align}
For $t\gtrsim t_c$, the linear-in-$t$ correction terms start to dominate over the $\sim 1/t$ term and $S_a + S_b$ becomes less than $S_c$, favoring the longer $ab$-path over the $c$-path on the hypotenuse. The existence of the time scale $t_c$ beyond which the edge-AB path is favored provides justification to the AB interpretation using the hole area and is the key assertion of our analysis. 

%\lee{Perhaps we can drop this next paragraph. It is a bit confusing to what happens when alpha=beta. Perhpas it is best not to discuss that case.} 
%A simple, geometric understanding of the crossover from the $c$-path to the $ab$-path, despite the longer length of the latter, can be given. First, let us assume $\alpha = \beta$ and  usinging $t' = t\times a/(a+b)$ and $t-t' = t \times  b/(a+b)$ for simplicity, one can compare the $\sim \alpha$ terms in $S_a + S_b$ with the similar term in $S_c$:
%
%\begin{align}
 %   {\rm from} ~ S_a + S_b & : \alpha \frac{a^3 + b^3}{a+b} t , \nn 
 %   {\rm from } ~ S_c  & : \alpha (a^2 + b^2) t . 
%\end{align}
%The first line in this equation is always smaller than the second line due to the identity $a^3 + b^3 < (a^2 + b^2) (a+b)$, and the difference can only grow with time $t$. Next we assume $\alpha=0$. for short time the $c$-path is preferred over the $ab$-path by virtue of another geometric identity $a^2 + b^2 < (a+b)^2$, but this advantage diminishes with time as $1/t$. At sufficiently long time $t$, the advantage of the $S_a + S_b$ coming from the other geometric identity $a^3 + b^3 < (a^2 + b^2) (a+b)$ takes over. 

\begin{figure}[tb]
\includegraphics[width=0.45\textwidth]{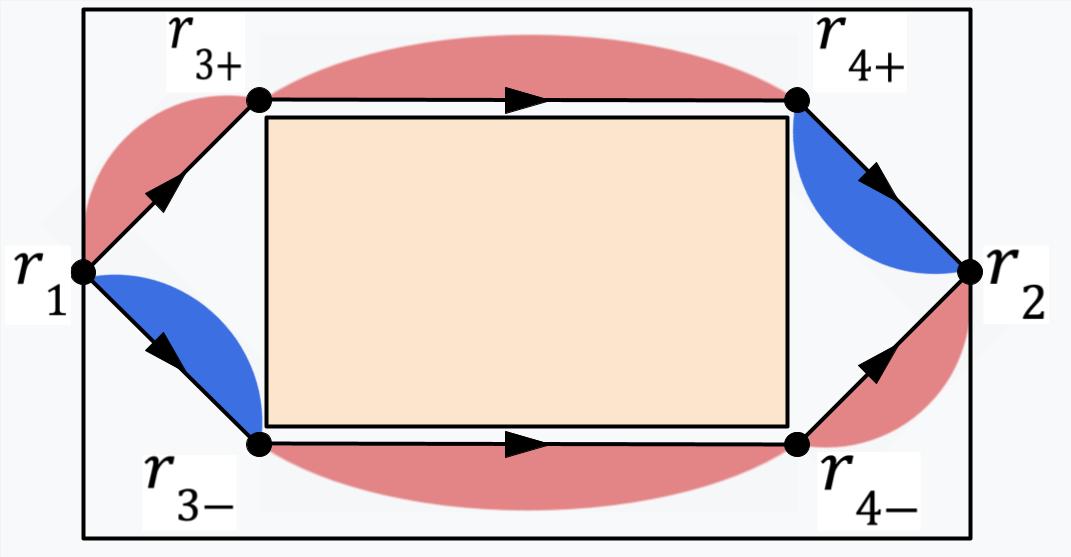}
\caption{Fluctuating paths give rise to area fluctuations. $\Delta A > 0$ ($\Delta A < 0$) fluctuations are shown as red (blue) areas.} 
\label{fig:areas}
\end{figure}

\section{Effective Action} 
\label{sec:effective-S} 

Now we give physical justification to the $\propto t$ term in the effective action, estimate the coefficients $\alpha, \beta$, and carve out the conditions under which the crossover of paths from short to long can take place. As schematically shown in Fig. \ref{fig:areas}, we consider fluctuations away from the classical paths and their effects on the action at the Gaussian level. A well-known correction to the classical action $S\sim (m^* /2) (l^2/t)$ is the prefactor $t^{-d/2}$ where $d$ is the spatial dimension. Such correction, though, will modify the analysis (which concerns quantities in the exponent) only by logarithmic amount and not play a vital role. A more pertinent correction comes from changes in the AB phase $\oint {\bf A} \cdot d{\bf r}$. As shown in Fig. \ref{fig:areas}, fluctuation of the classical path implies the change in the area, giving rise to the change in the phase factor 

\begin{align}
    \exp \left( \frac{2\pi i B}{\Phi^*_0} \Delta A \right) 
\end{align}
where $\Delta A$ is the area enclosed between the classical and the fluctuating (non-classical) path - red and blue region in Fig. \ref{fig:areas}. For the `free' region the area $\Delta A$ takes both positive and negative values. This is not the case, however, if the classical path is against the wall, i.e. it travels along the edges of the hole. In this case the area fluctuation is always of one sign: $\Delta A > 0$. For the two paths shown in Fig. \ref{fig:paths}(c), the difference arises in going from ${\bf r}_2$ to ${\bf r}_{3\pm}$ and from ${\bf r}_{4\pm}$ to ${\bf r}_2$. In both cases, one only needs to compare the effective actions for the hypotenuse and the base+perpendicular paths of Fig. \ref{fig:paths}(c). 

If the time $t$ is not too long, the typical fluctuating path will be `smooth' and unlikely to cross the classical path. In that case the area $\Delta A$ can be estimated reasonably well by the length $l$ of the classical path multiplied by the width in the transverse direction. Given that the governing equation (\ref{eq:diff-eq}) is a diffusion equation, the condition for this is $t<t_D$ where the diffusion time is given by 
\begin{align} t_D \sim l^2/D. \end{align} 
The size of the transverse fluctuation can be estimated by using the one-dimensional diffusion equation (as we are interested in the fluctuation in the transverse direction to the classical path), 
\begin{align}
\frac{\partial \psi}{\partial t} = \frac{\hbar}{2m^*} \frac{\partial^2 \psi}{\partial y_\perp^2}  
\equiv D \frac{\partial^2 \psi}{\partial y_\perp^2} ,
\end{align}
for the transverse coordinate $y_\perp$. Using its well-known solution, $\psi (y_\perp , t) =(4\pi D t)^{-1/2} \exp (-y_\perp^2 / 4Dt )$, $\langle y_\perp^2 \rangle = 2Dt$, we estimate 

\begin{align}
    \Delta y_\perp & \sim \sqrt{2Dt} = \sqrt{\hbar t/m^*}  = \xi \sqrt{2t/\tau} , \nn
    \Delta A & \sim l \xi \sqrt{2t/\tau} \label{eq:Delta-A} 
\end{align}
where $l$ is the length of a particular classical trajectory and $t$ is the time taken to traverse it. 

Evaluation of the two-particle correlator, Eq. (\ref{eq:2-ptl-propagator}), ultimately depends on summing over various paths, and this implies summation over various areas $\Delta A$. We therefore introduce the Gaussian 
distribution function for the area
\begin{align}
    P(\Delta A ) = \frac{e^{ - (\Delta A)^2 / 2\sigma_A^2 }}{\sqrt{2\pi \sigma_A^2}}  , 
    \left( \sigma_A = l \xi \sqrt{ \frac{2t}{\tau}} \right) ,
\end{align}
and sum over $\Delta A$ with this probability. For the area fluctuation against the edge where $\Delta A > 0$, the normalization factor is only half as large: $\sqrt{2\pi \sigma_A^2} \rightarrow \sqrt{2\pi \sigma_A^2}/2$. The following {\it exact} integral identities are useful in estimating the effect of area fluctuation both without and with the wall: 
\begin{widetext}
\begin{align}
\int_{-\infty}^\infty \exp\left( 2\pi i \frac{ B \Delta A }{\Phi^*_0 } \right) P(\Delta A) d (\Delta A) & =  e^{-\alpha l^2 t} \nn 
\int_{0}^\infty \exp\left( 2\pi i \frac{ B \Delta A }{\Phi^*_0 } \right) P(\Delta A) d (\Delta A) & = e^{-\alpha l^2 t} \left[ 1 + i {\rm erfi}( \sqrt{\alpha t} l ) \right] .  \label{eq:integral-identities} 
\end{align}
\end{widetext}
In the case of the first line, one can think of the $\alpha l^2 t$ in the exponent as an addition to the classical action. Comparing this to the effective action in Eq. (\ref{eq:Sabc0}), we realize that this $\alpha$ is precisely the phenomenological $\alpha$ there, but now it is explicitly given by
\begin{align} \alpha = ( 2\pi B \xi /\Phi^*_0 )^2 /\tau . \label{eq:alpha} \end{align} 
The imaginary error function erfi($x$) in the second line of the identity has the property
\begin{align}
    {\rm erfi}(x) \equiv \frac{2}{\sqrt{\pi}} \int_0^\infty e^{y^2} dy = e^{x^2} D(x) , 
\end{align}
and the {\it Dawson function} $D(x)$ has the asymptotic properties
\begin{align}
    D(x) & \sim x ~~~ (x \ll 1) \nn 
    D(x) & \sim 1/2x ~~~  (x \gg 1) , \label{eq:algebraic-Dawson}
\end{align}
with the maximum value less than $1$ at $x \approx O(1)$. Using the Dawson function one can re-write the half-infinite integral in the second line of Eq. (\ref{eq:integral-identities}) as
\begin{align} 
& \int_{0}^\infty \exp\left( 2\pi i \frac{ B \Delta A }{\Phi^*_0 } \right) P(\Delta A) d (\Delta A) \nn 
& ~~ = e^{-\alpha l^2 t} + i D ( \sqrt{\alpha t} l ) .  \label{eq:Dawson} 
\end{align} 
If we could ignore the Dawson function and keep the first term only, it means the parameter $\beta$ equals $\alpha$. If instead the Dawson function dominates at long times, the decay with time is only algebraic, and $\beta$ is effectively zero. The crossover from $\beta \approx \alpha$ to $\beta \approx 0$ takes place around the time $t_\beta$ which is estimated by $\alpha l^2 t_\beta \approx 1$, or in physical terms, 

\begin{align}
     \frac{t_\beta}{\tau} =\left(\frac{\Phi^*_0}{2\pi B \xi l } \right)^2 . \label{eq:t-estimate} 
\end{align}

We return to Eq. (\ref{eq:Sabc}) for $S_a + S_b - S_c$ and insert $\alpha$ from Eq. (\ref{eq:alpha}). We consider time $t>t_\beta$ (to be justified later) so that we can set  $\beta = 0$, to find
\begin{align}
S_a + S_b - S_c & \approx \frac{m^* ab}{t} \Bigl( 1 - \left[1 + \frac{b^2}{a (a+b)} \right] \left( \frac{t}{t_c} \right)^2 \nn 
& ~~ - \frac{a^4}{(a+b)^4} \left(  \frac{t}{t_c} \right)^4  \Bigr) .
\label{eq:Sabc2}
\end{align}
Accordingly, the crossover from the $c$-path to the $ab$-path takes place when 
\begin{align} \frac{t}{t_c}  \gtrsim \sqrt{\frac{a^2 + ab}{a^2 + ab + b^2}} .  \end{align} 
The factor on the r.h.s. is of order one for typical aspect ratios $a:b$, and we obtain that for $t \gtrsim t_c$ the path crossover indeed takes place. We also need the estimate of the crossover time $t_c$ in terms of the relaxation time $\tau$ and other physical parameters
\begin{align}
    \frac{t_c}{\tau} = \frac{\sqrt{m^*/\alpha}}{\tau} =  \frac{1}{2\sqrt{2}\pi} \frac{\Phi^*_0}{ B \xi^2} . \label{eq:t-crossover}  
\end{align}
Comparing with Eq. \eqref{eq:t-estimate}, we find the ratio

\begin{align}
    \frac{t_c}{t_\beta}  =  \sqrt{2}\pi \frac{ B l^2}{\Phi^*_0} . 
    \label{eq:ratio}  
\end{align}
Since $l$ is roughly half the linear dimension of the hole, we get $B l^2 \approx B A/4$, $A$ being the area of the hole. Thus for a field $B_1$ defined by $B_1 A=\Phi^*_0$ which corresponds to the first $AB$ oscillation, $t_c$ is already comparable or larger than $t_\beta$ and this ratio increases for larger $B$. This justifies our use of $\beta=0$.
%using the coherence $\xi = 2 \times 10^{-8}$m to reach the final expression. Inserting $B=B_{1}$ gives $t_c /\tau \approx 8$.} 

% The other two integrals for $A_{34}^\pm$ require that $\langle A_{34}^\pm \rangle = \Delta A_{34} = r_{34} l \xi \sqrt{2(t_4 - t_3) /\tau}$ be nonzro, as well as its Gaussian fluctuation being nonzero. This leads to the effective increase of the area in the AB phase as
% %
% \begin{align}
%     A_c \rightarrow A_c + 2 r_{34} l \xi \sqrt{2 (t_4 - t_3 )/\tau} . 
% \end{align}

The analysis so far only concerned a part of the overall trajectory going from ${\bf r}_1$ to ${\bf r}_2$. For a complete analysis let us go back the full edge-AB trajectory ${\bf r}_1$ to ${\bf r}_2$ and estimate its effective action: 
\begin{align}
S_{13} & = \frac{m^*}{2} \frac{(a+b)^2 }{t_3 \!- \!t_1}  + \alpha \frac{a^3}{a+b} (t_3 - t_1 ) ,  \nn 
%
%S_{34} & = \frac{m^*}{2} \frac{d^2}{t_4 \!-\! t_3}  + \alpha d^2 (t_4 - t_3 ) , \nn 
S_{34} & = \frac{m^*}{2} \frac{d^2}{t_4 \!-\! t_3} , \nn 
S_{42} & = \frac{m^*}{2} \frac{(a+b)^2 }{t_2 \!- \!t_4}  + \alpha \frac{a^3}{a+b} (t_2 - t_4 )  .
\label{eq:effective-action} 
\end{align}
The subscript in $S_{ij}$ means the trajectory from ${\bf r}_i$ to ${\bf r}_j$. Note that the $ab$-path is already assumed in writing down $S_{13}$ and $S_{42}$ above. The linear-$t$ term is absent in $S_{34}$ since the fluctuation there is against the wall and we already showed $\beta =0$ in that case. An estimate of $t_3, t_4$ minimizing the total action $S = S_{13} + S_{34} + S_{42}$ is given by
\begin{align} t_3 - t_1 & = t_2 -t_4 = \frac{a+b}{l_{\rm eff}} (t_2 - t_1) , \nn 
t_4 - t_3 & = \frac{d}{l_{\rm eff}} (t_2 - t_1)  ,  \label{eq:time-estimate} \end{align} 
where $l_{\rm eff} = 2(a + b) + d$. The total action is
\begin{align}
S& = S_{13} + S_{34} + S_{42} \nn 
& = \frac{m^*}{2} \frac{l_{\rm eff}^2}{t_2 \!-\! t_1} + \left( \alpha \frac{2 a^3}{l_{\rm eff}} + \frac{1}{\tau} \right)  (t_2 - t_1) , \label{eq:effective-total-S}
\end{align}
where we now include the exponential time factor $e^{-(t_2 - t_1) /\tau}$ as part of the effective action. 

The conductivity $\sigma ({\bf r}_1 , {\bf r}_2 )$ follows from taking the Fourier transform of the current-current correlation $\langle {\bf J} ({\bf r}_1 t_1 ) \cdot {\bf J} ( {\bf r}_2 t_2) \rangle$ with respect to $e^{-i\omega (t_2 - t_1)}$ and taking $\omega \rightarrow 0$ limit. In turn, $\langle {\bf J} ({\bf r}_1 t_1 ) \cdot {\bf J} ({\bf r}_2 t_2) \rangle$ is dominated by the path that maximizes the propagator, Eq. (\ref{eq:2-ptl-propagator}), or minimizes the total action $S$ in Eq. (\ref{eq:effective-total-S}). The minimization of $S$ with respect to $t_2 - t_1$ gives 
\begin{align}
    t_2 - t_1 = t^* = \frac{t_c }{\sqrt{2 \left[ 2 a^3/ l_{\rm eff}^3 + 1/\alpha \tau  l_{\rm eff}^2 \right] }}  \label{eq:t-star}
\end{align}
where $t_c$ is the crossover time in Eq. (\ref{eq:t-crossover}). Among the two terms in the denominator, 
\begin{align}
    \frac{1}{\alpha \tau l_{\rm eff}^2} = \left( \frac{\Phi^*_0}{2 \pi B \xi l_{\rm eff} } \right)^2  \approx \left( \frac{B_1}{B} \right)^2 \left( \frac{A}{2 \pi  \xi l_{\rm eff} } \right)^2 . 
\end{align}
We estimate $\xi = 2\times 10^{-8}$m and  $l_{\rm eff} = 5\times 10^{-7}$m. By way of comparison, the area of the hole $A$ in the experiment is estimated to be $2.76 \times 10^{-14}$m$^2$ , which corresponds to a linear dimension of about $1.7 \times 10^{-7}$m. Thus $l_{\rm eff}$ is about 3 times the dimension of the hole. With the $2\pi$ factor in our favor, we find that this expression is comparable or less than one already at $B=B_1$, the field of the first $AB$ oscillation and will decrease rapidly at larger magnetic field. On the other hand,   the first term in the denominator of Eq. (\ref{eq:t-star}) is independent of $B$. For this reason we keep only the first term in the denominator and write
\begin{align}
%    t^* \approx t_c \sqrt{\frac{l_{\rm eff}^3}{2 (2a^3 + d^3 )}} .
    t^* \approx t_c \sqrt{\frac{l_{\rm eff}^3}{4 a^3}} .
\end{align}

The value of $t^*$ is then used to estimate the time to go from ${\bf r}_1$ to ${\bf r}_{3\pm}$ as 
\begin{align}
    t_3 -t_1 = \frac{a+b}{l_{\rm eff}} t^* =  t_c \sqrt{\frac{(2a + 2b + d)(a+b)^2}{4a^3}} . \label{eq:estimate-of-t3t1} 
\end{align}
Recall, as we argued earlier, that the time $t_3 - t_1$ to complete the $ab$-path needs to be larger than $t_c$ in order to justify the crossover from $c$-path to the $ab$-path. Now let's estimate $t_3 - t_1$ explicitly. For $a:b:d=1:1:1$, for instance, we get $t_3 - t_1 = \sqrt{5}t_c > t_c$. For $a:b:d=1:1:2$ we get an even larger $t_3 - t_1 = \sqrt{6} t_c$. The current device setup~\cite{6e-wang} is therefore consistent with the statement that {\it the time to travel between ${\bf r}_1$ and ${\bf r}_{3\pm}$ is sufficiently long to justify the choice of $ab$-path over the $c$-path}. Note further than this condition is geometric, i.e. does not depend on the strength of the magnetic field $B$.

% \han{One can repeat the calculations of Eqs. (\ref{eq:effective-action}) through (\ref{eq:estimate-of-t3t1}) assuming $\beta = \alpha$ instead of $\beta = 0$. The segment-wise actions of (\ref{eq:effective-action}) become
% %
% \begin{align}
% S_{13} & = \frac{m^*}{2} \frac{(a+b)^2 }{t_3 \!- \!t_1}  + \alpha (a^2 +b^2 - ab) (t_3 - t_1 ) ,  \nn 
% %
% S_{34} & = \frac{m^*}{2} \frac{d^2}{t_4 \!-\! t_3}  + \alpha d^2 (t_4 - t_3 ) , \nn  
% %
% S_{42} & = \frac{m^*}{2} \frac{(a+b)^2 }{t_2 \!- \!t_4}  + \alpha (a^2 +b^2 - ab) (t_2 - t_4 )  .
% \label{eq:effective-action2} 
% \end{align}
% %
% The estimate of $t_3, t_4$ minimizing the total action is the same as Eq. (\ref{eq:time-estimate}). The total action is
% %
% \begin{align}
% S& = \frac{m^*}{2} \frac{l_{\rm eff}^2}{t_2 \!-\! t_1} \nn 
% %
% & + \left( \alpha \frac{2 a^3 + 2b^3 + d^3}{l_{\rm eff}} + \frac{1}{\tau} \right)  (t_2 - t_1) , \label{eq:effective-total-S}
% \end{align}
% %
% where one can see the only difference from the previous expression (\ref{eq:effective-total-S}) in the replacement $2a^3 \rightarrow 2a^3 + 2b^3 + b^3$. This changes the estimate of $t_3 - t_1$ in Eq. (\ref{eq:estimate-of-t3t1}) to
% %
% \begin{align}
%     t_3 -t_1 = t_c \sqrt{\frac{(2a + 2b + d)(a+b)^2}{2( 2a^3 + 2b^3 + d^3 )}} . \label{eq:estimate-of-t3t1} 
% \end{align}
% %
% For the aspect ratio $a:b:d= 1:1:1$ we get $t_3 - t_1 = \sqrt{2} t_c$, and with $a:b:d=1:1:2$, $t_3 - t_1 = t_c$. In all, the condition for the $c$- to $ab$-path crossover is not changed qualitatively whether we choose $\beta =0$ or $\beta=\alpha$.}

For the $b$- and $d$-trajectories, we already concluded that there are no corrections to the classical action because $\beta$ is effective zero. However, the area covered by a given trajectory always satisfies $\Delta A >0$ and this might give rise to the increase in the effective area from the geometric area of the hole, $2b\times d$. Contrary to the naive expectation, this is not the case. This can be seen easily from the expression in Eq. (\ref{eq:Dawson}), which contains no oscillatory terms. To be more explicit, for the segments of the path hugging the edge, we may use the half-integral identities in Eqs. (\ref{eq:Dawson}) and (\ref{eq:algebraic-Dawson}) to obtain the compound action
\begin{align}
& \left( \frac{i}{2\sqrt{\alpha (t_3 \!-\! t_1)} b} \right)^2 \left( \frac{i}{2\sqrt{\alpha (t_4 \!-\! t_3)} d} \right)^2 \left( \frac{i}{2\sqrt{\alpha (t_2 \!-\! t_4)} b} \right)^2 \nn 
& \sim - \frac{1}{(t_2 - t_1)^3} .  
\end{align}
This factor, being purely real, will have no contribution to the AB phase $\exp [ (2\pi i/\Phi^*_0) \oint {\bf A}\cdot d{\bf r} ]$. Thus, contrary to naive expectation, the effective area stays the same as the geometric area of the inner square: $A_{\rm eff} = 2b \times d$. The situation is analogous to what happens in three dimensional quantum oscillations, where the oscillation period is given by the extremal area, despite a broad distributions in the cross-section area that deviates from the extremal area with the same sign. Mathematically this is because the integral in the second line of Eq. (\ref{eq:integral-identities} has a sharp cutoff and the effective area is determined by that cut-off.

\section{Discussion} 
Several time scales were introduced and discussed throughout the paper. Among them, the diffusion time $t_D$ was crucial in allowing us to use the Gaussian probability function $P(\Delta A)$ for the area fluctuation that increases diffusively with time $\Delta A \propto \sqrt{t}$, Eq. (\ref{eq:Delta-A}). %This kind of picture holds when $t$ is less than the diffusion time $t_D \sim l^2/D$, otherwise the boundary of $\Delta A$ starts to wiggle too much and the simple diffusion relation $\Delta A \sim \sqrt{t}$ no longer applies. It is vital, then, that $t_c$ be less than $t_D$. 
Since $D=\hbar/2m^*$ and $m^* = \hbar \tau / 2\xi^2$, 
\begin{align} t_D & = l^2 /D = (l/\xi)^2 \tau %, \nn 
%t_D / t_c & = 2\sqrt{2} \pi (B l^2 /\Phi^*_0 ) . 
\end{align} 
%
%For typical field $B$, we are safely in the $t_c < t_D$ regime and the use of the diffusion-type kernel for the area function is justified. 
The other two time scales, $t_\beta$ and $t_c$, were defined in Eq. (\ref{eq:t-estimate}) and (\ref{eq:t-crossover}), respectively. Putting all three time scales together gives 
%Another time scale $t_\beta$ of concern is the crossover from $\beta \approx \alpha$ to $\beta \approx 0$. As discussed earlier in Eq. (\ref{eq:t-estimate}), $t_\beta = \tau \times (\Phi^*_0 /\Phi )^2$ where $\Phi$ is comparable to the net flux through the device. In fact, there is a nice succession of ratios among the three time scales,  
%
\begin{align}
\frac{t_\beta}{\tau} : \frac{t_c}{\tau} : \frac{t_D}{\tau} & \approx \left(\frac{\Phi^*_0}{2\pi B \xi l} \right)^2 : \frac{\Phi^*_0}{ 2\sqrt{2} \pi  B \xi^2} : \left(\frac{l}{\xi} \right)^2 \nn 
& \approx \left(\frac{\Phi^*_0}{2\pi B l^2} \right)^2 : \frac{\Phi^*_0}{ 2\sqrt{2} \pi  B l^2 }  : 1 . 
\end{align}
For instance, the condition $t_D \gtrsim t_c$ amounts to having $1 \gtrsim \Phi^*_0 / ( 2\sqrt{2} \pi  B l^2 )$, which is nicely fulfilled if we take $B l^2$ to be the AB flux through the hole, so that $B \gtrsim B_1$ , i.e. at least one $AB$ oscillations have been seen. The inequality $t_c \gtrsim t_\beta$ holds under the similar condition.

As mentioned in the introduction, this paper does not address the microscopic justification of the charge-$6e$ formation in the CVS superconductor. We take it for granted and write down the Ginzburg-Landau theory accordingly. The starting point of our investigation is the observation that the experimental setup of Ref. \onlinecite{6e-wang} calls for a theoretical justification of the effective area one should use in estimating the flux quantum. The effective area used by the authors of the experiment suggests that the area of hole (the inner square) is playing a dominant role, but on the basis of some earlier literature, \cite{kogan01,kogan04} it is not clear whether this is justified.  Our investigation shows that the use of the extremal area - the area of the inner square - does get justified under the experimentally viable circumstances $\Phi/ \Phi^*_0 \gtrsim 1$ ($\Phi$=flux through the hole) if the pairing order is fluctuating and has only short range order. 

Heuristically, the preferred path that dominate the contribution to the correlation function is not always the geometrically preferrred path, but the one that likes to `stick to the edge' because there the fluctuation (and hence the effective action) is less. This simple picture arising from our analysis is obviously applicable to the more mundane, charge-$2e$ superconductors under the similar device setup, as long as we are in the fluctuation regime. We surmise that this result may be more general, as long as the system is described by a sum over diffusing Feynman paths. The basic principle rests on the analog of Eq. \eqref{eq:integral-identities} which states that the average over a distribution of $AB$  areas is dominated by the extremal area in the presence of a sharp cut-off, and not by the average area. Thus the result that the $AB$ oscillation is given by the area of the hole may also apply to the charge-$e$ transport in a normal metal in the thick-rim geometry.

\acknowledgments
J.H.H. was supported by  NRF-2019R1A6A1A10073079. He also acknowledges financial support from EPIQS Moore theory centers at MIT and Harvard, and thanks Manhyung Han for preparing the figures. Hospitality extended by IBS-CCES at Seoul National University during the completion of this work is gratefully acknowledged. P.L. acknowledges the support by DOE office of Basic Sciences Grant No. DE-FG02-03ER46076. We thank Ziqiang Wang for enlightening discussion and comments on the manuscript. 

\bibliography{6e}
\end{document}